\documentstyle[prl,twocolumn,epsf,aps]{revtex}

\begin{document}
\draft
\title{Spin-orbit coupling and electron spin resonance
 theory for carbon nanotubes}
\author{A. De Martino,$^{1}$ R. Egger,$^{1}$ 
K. Hallberg,$^{2}$ and C.A. Balseiro$^2$}
\address{${}^{1}$~Institut f\"ur Theoretische Physik, 
Heinrich-Heine Universit\"at,
D-40225 D\"usseldorf, Germany\\
${}^{2}$~Instituto Balseiro, Centro Atomico,
8400 S.C. de Bariloche, Argentina}
\date{Date: \today}
\maketitle
\begin{abstract}
A theoretical description of electron spin resonance (ESR)
in 1D interacting metals is given, with primary emphasis on
carbon nanotubes.  The spin-orbit coupling is derived, and the 
resulting ESR spectrum is analyzed using a low-energy field theory.
Drastic differences in the ESR spectra of
single-wall and multi-wall nanotubes are found.  For single-wall tubes,
the predicted double peak spectrum is linked to spin-charge separation.
For multi-wall tubes,  a single narrow asymmetric peak is expected.
\end{abstract}

\narrowtext

Electron spin resonance (ESR) serves as a valuable tool
to experimentally probe the intrinsic spin dynamics of many systems.
In ESR experiments one applies a static magnetic field
and measures the absorption of radiation polarized
perpendicular to the field direction. In the absence of 
$SU(2)$ spin symmetry breaking terms in the system Hamiltonian,
the absorption intensity is then simply a $\delta$-peak 
at the Zeeman energy \cite{OA}.
Since spin-orbit (SO) interactions are generally the
leading terms breaking the $SU(2)$ invariance, 
deviations in the ESR intensity from the $\delta$-peak, e.g.~shifts or
broadenings, are directly connected to these couplings.
In this Letter we theoretically address the spin-orbit interaction and the
resulting ESR spectrum for interacting 1D metallic conductors, 
in particular for carbon nanotubes.
Nanotubes constitute a new class of mesoscopic quantum wires characterized
by the interplay of strong electron-electron interactions,
reduced dimensionality, disorder, and unconventional spin dynamics
\cite{dekker99,ll1,forro,egger97,balents}.  ESR is an important
technique to elucidate aspects of this interplay inaccessible
to (charge) transport experiments.  For interacting many-body systems,
surprisingly little is known about ESR although it represents
an interesting theoretical problem. 

Two main classes of nanotubes may be distinguished,
namely single-wall nanotubes (SWNTs) which consist
of just one wrapped-up graphite sheet, and
multi-wall nanotubes (MWNTs).  MWNTs contain additional inner shells, 
but transport is generally limited to the outermost shell \cite{forro}.
Evidence for the Luttinger liquid (LL) behavior of interacting
1D electrons has been reported for charge transport 
in SWNTs \cite{ll1}, where one also
expects to find spin-charge separation \cite{egger97,balents}.
Conventional wisdom holds that the SO coupling in 
1D conductors destroys spin-charge separation \cite{moroz}.  
Below we show that this statement is {\sl incorrect}.  
Indeed, the SO interaction considered
in Ref.~\cite{moroz} was intended for the limited class of  
semiconductor quantum wires in strong Rashba and 
confinement electric fields, but in fact does not represent
the generic SO Hamiltonian for 1D conductors.
The latter is derived below and determines the ESR intensity in
SWNTs and MWNTs.  A totally different ESR spectrum 
compared to expectations based on Ref.~\cite{moroz} emerges.
In particular, the single $\delta$-peak is split
into {\sl two} narrow peaks in SWNTs, 
while for the SO coupling of Ref.~\cite{moroz}
the spectrum forms
a broad band with thresholds at the lower and upper edge 
\cite{esrqw}.  This qualitative difference
can be traced back to the fact that the SO interaction in SWNTs
[see Eq.~(\ref{contHam}) below]
does {\sl not}\, spoil spin-charge separation.
Experimental observation of the peak splitting could
therefore provide strong evidence for the elusive
phenomenon of spin-charge separation.
In MWNTs, inner shells cause a rather strong 
Rashba-type SO coupling, leading instead to a {\sl single} narrow asymmetric
ESR peak.
To experimentally observe our predictions,
it will be crucial to work with samples free of magnetic impurities
whose presence has drastically affected previous
ESR measurements for nanotubes \cite{forro}.

In the standard Faraday configuration,
the ESR intensity at frequency $\omega$ is 
proportional to the Fourier transform of the transverse spin-spin 
correlation function \cite{OA},
\begin{equation} \label{int1}
I(\omega)= \int \! dt \, e^{i\omega t} \langle S^+(t) S^-(0) \rangle ,
\end{equation}
where the static magnetic field
points along the $z$-axis and
$\vec S = \sum_i\vec S_i $ is the total spin operator. 
Equation (\ref{int1}) is connected to the dynamical susceptibility via
$I(\omega)\sim \omega \chi^{\prime\prime}(\omega)$.
The Hamiltonian is  $H=H_0 + H_Z + H'$,
where $H_0$ represents the $SU(2)$ invariant nanotube Hamiltonian
including electron-electron interactions, 
$H_{Z} = -BS^z$ is the Zeeman term \cite{foot}, and $H'$ represents
$SU(2)$ breaking terms, in particular the SO coupling. 
Inserting a complete set of eigenstates $|a \rangle$ of $H$ in 
Eq.~(\ref{int1}), the ESR intensity  follows as
\[
I(\omega) = \frac{1}{Z} \sum_{a,b} e^{- E_b/T} 
\delta(\omega -(E_a-E_b)) \,
| \langle a | S^- | b \rangle |^2.
\]
If $H$ is $SU(2)$ invariant (apart from $H_Z$), $I(\omega)$ only receives 
contributions from matrix elements between 
eigenstates with equal total spin $S_a= S_b$.  Then 
all states with $S^z_a=S^z_b-1$ will contribute to form a 
$\delta$-peak at frequency $\omega=B$.
At zero temperature,  the application of a magnetic field $B$, 
taken as large enough to overcome a spin gap possibly present at $B=0$, 
leads to a ground state with finite magnetization, $S_0 \neq 0$, and 
the states with $S^z_a=S^z_0-1$  yield the
$\delta$-peak. Any perturbation preserving $SU(2)$ invariance 
will neither shift nor broaden this peak, even
at finite temperature, and it
is therefore crucial to identify $H'$. 
For quantum spin chains, staggered magnetic fields and 
Dzyaloshinskii-Moriya interactions  have been emphasized \cite{OA}.

Let us start with the derivation of the SO interaction in nanotubes. 
In this derivation we neglect electron-electron interactions 
which only weakly renormalize the SO strength \cite{chenraikh}.
In a single-particle picture, the SO interaction then appears because an 
electron moving in the electrostatic potential 
$\Phi(\vec r)$, e.g.~due to the ions, sees an effective magnetic field 
$\vec v \times \nabla \Phi$. With $\vec p = m \vec v$
and the standard Pauli matrices $\vec\sigma$, the SO interaction
reads in second-quantized form: 
\begin{equation} \label{SOham2}
H' = -\frac{g_e \mu_B}{4m} \int d\vec{r} \, \, \Psi^\dagger \left[ 
(\vec p \times \nabla \Phi) \cdot \vec \sigma \right] \Psi.
\end{equation}
The electron spinor field
$\Psi_\sigma(\vec r)$, defined on the wrapped graphite 
sheet, can be expressed in terms of the electron operators $c_i$ 
for honeycomb lattice site $i$ at $\vec r_i$,
$\Psi_\sigma (\vec r) = \sum_i \chi(\vec r - \vec r_i) \ c_{i\sigma}$, 
where $\chi(\vec r)$ is the $2 p_z$ orbital wavefunction.
The localized orbitals can be chosen 
as real-valued functions even when hybridization with $2s$ 
orbitals is taken into account,
but their specific form is of no immediate interest here.
We then obtain the SO interaction, see also Refs.~\cite{bonesteel,ando},
\begin{equation} \label{SOham3}
H' = \sum_{\langle jk \rangle} \,\,
 i c^\dagger_j (\vec u_{jk} \cdot \vec \sigma) c^{}_k  + \,{\rm h.c.}  
\end{equation}
which indeed breaks $SU(2)$ symmetry.   With obvious modifications,
Eq.~(\ref{SOham3}) applies to other 1D conductors
and thus represents a generic SO Hamiltonian. 
The SO vector $\vec{u}_{jk}=-\vec{u}_{kj}$ has real-valued entries,
\begin{equation}\label{sovect}
\vec u_{jk} = \frac{g_e\mu_B}{4m} \int d\vec r \, \Phi(\vec r)
\left[ \nabla \chi (\vec r - \vec r_j) \times 
\nabla \chi (\vec r - \vec r_k)  \right]  .
\end{equation} 
The on-site term ($j=k$) is identically zero, and
since the overlap decreases exponentially with $|\vec r_j - \vec r_k|$, 
we keep only nearest-neighbor terms in Eq.~(\ref{SOham3}). 
We mention in passing that Eq.~(\ref{SOham3}) has 
previously been found from $\vec k \cdot \vec p\,$
theory by Ando \cite{ando}.  However, his approach makes rather special
model assumptions and is technically demanding, yet it
does not allow to reliably compute the SO vector $\vec{u}_{jk}$ to 
better accuracy than specified in Eq.~(\ref{sovect}).  
In addition, the effect of SO interactions
within the low-energy theory of nanotubes \cite{egger97} 
has not been analyzed.  We therefore take Eq.~(\ref{SOham3}) as
the SO Hamiltonian for SWNTs and MWNTs, with the SO vector 
(\ref{sovect}).  This formulation also allows to incorporate
electric fields due to impurities or close-by electrodes
in a simple and elegant manner.

Let us first turn to SWNTs, where  SO couplings
are expected to be small. This can be rationalized from our approach
since the SO vector (\ref{sovect}) {\sl vanishes} by symmetry
for an ideal 2D honeycomb lattice.  A finite (nearest-neighbor)
SO coupling can only arise due to the curvature of the wrapped sheet,
stray fields from nearby gates, or due to defects, 
all of which break the high symmetry and in principle allow for
significant  SO couplings \cite{balents}.
Focusing on the curvature-induced SO coupling for nonchiral
SWNTs, the
SO vector only depends on bond direction,
$\vec u_{\vec r_i, \vec r_i+\vec \delta_a} = \vec u_a$, 
for the nearest-neighbor bonds $\vec\delta_a$ ($a=1,2,3$) of
the graphite sheet \cite{dekker99}.  

To make progress, we employ the effective 
field theory approach \cite{egger97}.
Neglecting the (here inessential) ``flavor'' index
due to the presence of two Fermi (K) points \cite{dekker99},
$H_0$ then corresponds to a spin-$1/2$ LL described by charge/spin
interaction parameters $K_c<1, K_s$ and 
velocities $v_{c/s}=v_F/K_{c/s}$ with the
Fermi velocity $v_F=8\times 10^5$ m/sec;
$SU(2)$ invariance of $H_0$ fixes $K_s=1$ \cite{GNT}.  
The LL Hamiltonian completely decouples when expressed 
in terms of spin and charge boson fields \cite{GNT},
and the Zeeman term $H_Z$ only affects the spin sector. 
Written in terms of right- and left-moving fermions $\psi_{R/L}$,
the continuum version of Eq.~(\ref{SOham3}) is 
(up to irrelevant terms) $H'=H_1+H_2$ with
\begin{eqnarray} 
&& H_1 = \int dx  \ \vec \lambda \cdot ( \vec J_L - \vec J_R),\label{contHam} 
\\ && H_2 =   \int dx 
\sum_{r=R/L} \psi^\dagger_r \,\vec \lambda'\cdot \vec 
\sigma \, i \partial_x \psi_r  + \mbox{h.c.}  \label{contHam2}
\end{eqnarray}
With the unit vector $\hat{e}_t$  
along the tube axis, we use
\[
\vec \lambda =  2\ \mbox{Im} \sum_a e^{i\vec K \cdot \vec\delta_a} 
\vec u_a, \quad \vec \lambda' = \sum_a e^{i\vec K \cdot\vec \delta_a}
(\hat{e}_t \cdot \vec \delta_a) \vec u_a \;,
\]
and neglect oscillatory terms  which average out on large length scales. 
These oscillations are governed by the wavevector $2k_F$ 
corresponding to the doping level $\mu$, 
$k_F=|\mu|/v_F$, with typical values $|\mu|\approx 0.3$ to 0.5~eV \cite{lemay}.
Finally, $\vec J_{R,L} = \psi_{R/L}^\dagger
(\vec\sigma/2) \psi_{R/L}^{}$ are the standard $SU(2)$ spin currents. 
The perturbation (\ref{contHam}) has scaling dimension $1$ (relevant)
whereas Eq.~(\ref{contHam2}) has dimension $2$ (marginal).  
Therefore the leading SO contribution retained in what follows is
Eq.~(\ref{contHam}).

Remarkably, the SO interaction (\ref{contHam}) acts exclusively 
in the spin sector and hence
does {\sl not spoil spin-charge separation}. 
As a consequence, since electron-electron interactions
affect only the charge sector,
the ESR intensity can be computed using
the equivalent fermionic spin Hamiltonian 
\begin{equation} \label{fermHam}
H_f = \sum_{r=R/L=\pm}  \int \ dx \left [-ir v_s
 \psi_{r}^\dagger \partial_x \psi^{}_{r}
+ \vec\lambda_r \cdot \vec{J}_r \right ],
\end{equation}
where $\vec\lambda_\pm = \vec B \pm \vec \lambda$.
Since $H_f$ is bilinear in the fermions, 
after some straightforward algebra,
the exact ESR spectrum follows for arbitrary temperature
\begin{equation} \label{swnt}
I(\omega)=\sum_{r=\pm} \left( 1+ \frac{\lambda^z_{r}}{\lambda_r} 
\right)^2  \frac{\lambda_r}{4v_s(1-e^{-\lambda_r/T})} \,
\delta(\omega-\lambda_{r}) , 
\end{equation}
with $\lambda_\pm=|\vec\lambda_\pm|$.
As a consequence of SO coupling, 
the single $\delta$-peak {\sl splits into two peaks} 
but there is no broadening.
The peak separation is $|\lambda_+ - \lambda_-|$, and
the peak heights are generally different.  To lowest order in
$\lambda/B$, the two peaks are located symmetrically around $\omega=B$.
Notice that for 
$\vec B$ perpendicular to the effective SO vector $\vec \lambda$,
the splitting is zero.  
It should be stressed that these results hold both for
the non-interacting and the interacting case.  
The double peak is therefore not directly related to the doubling
of singularities in the single-electron Greens function
commonly associated with spin-charge separation.
However, for the interacting case realized in SWNTs
\cite{ll1}, the double peak structure  is only possible if
spin-charge separation is present. Otherwise the
charge sector will mix in, leading to broad bands with 
threshold behaviors \cite{esrqw}.
Closer inspection shows that  inclusion of the subleading term 
(\ref{contHam2}) preserves the splitting into two peaks,
but the peaks now acquire a finite width $\sim |\vec \lambda'|$. 

Experimental observation of the predicted double-peak spectrum
could then provide strong evidence for spin-charge separation.
In practice, to get measurable intensities, one will have to 
work with an ensemble of SWNTs. The proposed experiment may be 
possible using electric-field-aligned SWNTs, or by employing 
arrays of identical SWNTS \cite{ibm}.
In more conventional samples containing many SWNTs, however, the 
SO vector $\vec \lambda$ can take any direction.
Assuming a uniform probability distribution for 
the orientation of $\vec\lambda$,  the average can
easily be done. For $T=0$ and $\vec\lambda'=0$, we find
\begin{equation}
\overline{I(\omega)} = 
\frac{[(\omega+B)^2-\lambda^2]^2}{16v_sB^3\lambda}
\theta(B+\lambda-\omega)\,\theta(\omega-|B-\lambda|)
\end{equation}
with the Heaviside function $\theta(z)$.
For $B>\lambda$, this asymmetric spectrum has  the width 
$\Delta \omega= 2\lambda$, which in turn allows to extract
the SO coupling strength $\lambda$ from ESR measurements.

For the remainder, we then focus on MWNTs where
we first contemplate a simple two-shell model.
Experimental evidence \cite{forro}
is consistent with the assumption that inter-shell tunneling
is strongly suppressed. Therefore doping due to charge transfer from
the substrate, the attached leads, or due to oxygen absorption,
should only affect the outermost shell, while 
$\mu\approx 0$ for inner shells.  Under this assumption,
basically all conduction electrons contributing to the ESR
signal reside in the outermost shell.
Moreover, the electrostatic potential $\Phi_o$ of the outer shell
differs from the inner-shell potential $\Phi_i$.
In effect, we can then restrict attention to
the outermost shell (of radius $R$) alone, but in  
a radial electric field of size $E\approx 2 c_{12}\Delta\Phi/R$,
where $c_{12}$ is the inter-shell capacitance per length
and $\Delta \Phi = \Phi_i-\Phi_o$.
The general expression (\ref{sovect})  for the SO vector 
then yields after some algebra for a given bond $\vec \delta_a$
\begin{equation} \label{uadef}
\vec u_a =  (u/v_F) \hat E \times \vec \delta_a,
\end{equation}
where $\hat E$ is a unit vector perpendicular to the tube
surface, and $u \approx  c_{12} e \Delta \Phi/(m^2 d R)$
with the C-C distance $d=1.42$~\AA.

To proceed, we turn to the low-energy theory, again 
for only one $K$ point.  The influence of interactions is
expected to be less dramatic in MWNTs compared to SWNTs,
and here we focus on the non-interacting case.  In real space,
we then have a Dirac Hamiltonian,
$H_0 =-i v_F \int \! d\vec r \, \Psi^\dagger  (\vec \tau \cdot \vec 
\nabla) \Psi$,  
where the integral extends over the tube surface and the
Pauli matrices $\vec\tau$ act in sublattice space.
The SO contribution appropriate for MWNTs follows by
inserting Eq.~(\ref{uadef}) into Eq.~(\ref{SOham3}), which yields
the manifestly Hermitian term
\begin{equation}
H' = -i u v_F \int \! d\vec r \ \ \Psi^\dagger 
 ( \tau^-\sigma^+ - \tau^+\sigma^- ) \Psi .
\end{equation}
We take the magnetic field parallel to the tube axis 
and include both orbital and Zeeman contributions \cite{foot2}. 

The full Hamiltonian can then be
diagonalized.  The dispersion relation contains four branches, 
$\epsilon(\vec k) = \pm \epsilon_\pm(k)$,  where the
$\pm$ signs are independent and
\begin{eqnarray} \label{disprel}
\epsilon_\pm(k)&=& \Bigl [ v_F^2 (k^2 + Q^2)+(B/2)^2 
 \pm v_F  \\ \nonumber
&\times&  \left \{( B^2 + 2v_F^2 Q^2)  k^2  + v_F^2 Q^4 \right \}^{1/2}
\Bigr ]^{1/2} , 
\end{eqnarray}
with $Q = 2\sqrt{2}u$  measuring the SO strength and $k=|\vec k|$.
Using Eq.~(\ref{disprel}), 
the  ESR intensity at finite temperature $T$ reads
\begin{eqnarray} \label{ESRmwnt}
&&I(\omega)= \frac{-1}{1-e^{-\omega/T}}
\int  \frac{d\vec k}{8 \pi} \frac{\Pi(\epsilon_+, k;\omega) \, 
\delta(\omega - \epsilon_+ + \epsilon_-)}
{\epsilon_+\epsilon_-(\epsilon^2_+ -\epsilon^2_-)^2} 
\times \nonumber \\
&&\left[n_F(\epsilon_+) - n_F(-\epsilon_+) - n_F(\epsilon_-) + 
n_F(-\epsilon_-)\right] ~,
\end{eqnarray}
where $n_F(\epsilon)= 1/(1+e^{-(\mu - \epsilon)/T})$ is the 
Fermi-Dirac distribution function.
The integral in Eq.~(\ref{ESRmwnt}) includes an integration
over the momentum parallel to the MWNT axis and 
a discrete summation over the quantized transverse momenta,
$k_\perp= (n-\phi)/R$, where $\phi = \pi R^2 B/ (h/e)$ is 
due to the orbital effect of the applied magnetic field 
and the summation extends over integer $n$.  Furthermore,
with $\bar \omega'= \omega' - \omega$ and
$f(\omega',k)=v_F^2 k^2 - (B/2 + \omega')^2$,
\begin{eqnarray}
&&\Pi(\omega', k;\omega)=  2f(\omega', k)
f(-\bar\omega',k) \times \nonumber\\
&&\left[(B/2 -\omega')(B/2+\bar\omega')-
v_F^2 k^2\right] \nonumber\\
&& - 2 v_F^2 Q^2 
\left[(B/2-\omega')(B/2-\bar\omega')
f(\omega', k)  \right. \nonumber \\
&&\left. + (B/2+\omega')(B/2+\bar\omega')
f(-\bar \omega',k) \right] .
\end{eqnarray}
As a simple check, for $Q=0$ one recovers the 
expected $\delta$-peak from Eq.~(\ref{ESRmwnt}).

The result (\ref{ESRmwnt}) can be 
understood in simple physical terms.
The ESR intensity receives contributions from transitions
between states of energy $\epsilon_-$ to states of energy $\epsilon_+$, 
and each contribution is weighted by a factor which takes into account the 
occupation of the levels. 
In order to arrive at Eq.~(\ref{ESRmwnt}), we have
neglected a contribution coming from transitions between
$-\epsilon_\pm$ and $\epsilon_\mp$ states. 
This is consistent because these transitions only contribute 
for very large frequencies $\omega \geq |\mu|$, while the 
frequency scales relevant to ESR are much lower. 
In addition, the signal given by this contribution is 
very small in comparison to the term that we keep.

\begin{figure} 
\epsfysize=8cm 
\epsffile{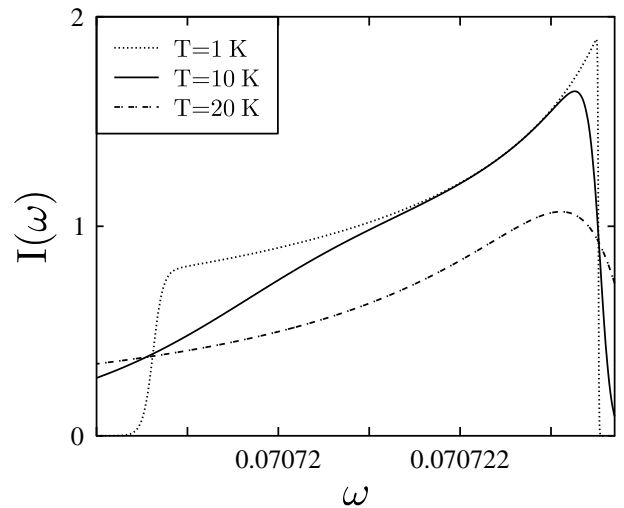} 
\caption{ \label{fig1} 
Typical ESR intensity of a MWNT at low temperatures. Parameters in the
plot are $\phi=0$, $\mu=0.1$, $B=0.0014$ and $v_F Q=0.05$, where
energies are measured in units of $2\pi \hbar v_F/R$.  The value of $B$
corresponds to a field of $10$~Tesla.  
Note the frequency units, pointing to a very narrow ESR peak. 
}
\end{figure}

Inspection of Eq.~(\ref{ESRmwnt}) shows that the
ESR spectrum of a MWNT at low temperatures
contains only a {\sl single narrow asymmetric peak}, 
whereas more structure  appears at higher 
temperature due to the activation of the transverse subbands \cite{long}. 
Here we focus on the low-temperature ESR spectrum,
which is shown in Fig.~\ref{fig1} for typical parameters.
The peak has an asymmetric lineshape which strongly depends
on temperature. At zero temperature, the intensity maximum is 
at the frequency
\begin{equation} \label{max}
\omega_0= \epsilon_Z \left[1 - \frac{\epsilon_Z^2}{2\mu^2}
- \frac{v_F^2 Q^2 B^2}{ 4\mu^2\epsilon_Z^2} + {\cal O}\left(
(\epsilon_Z/\mu)^3\right) \right]
\end{equation}
where $\epsilon_Z=\sqrt{B^2+2v_F^2 Q^2}$.
As is apparent in Fig.~\ref{fig1}, when increasing the temperature, 
the position of the maximum slowly moves to smaller frequencies.
The linewidth is of the order of 
\begin{equation}
\Delta \omega \approx v_F^2Q^2 B^2/\mu^3 \;.
\end{equation}

To conclude, we have presented a theoretical description of
the spin-orbit coupling and the ESR spectrum for 1D conductors,
in particular for carbon nanotubes.  
In SWNTs, spin-charge separation should not be affected by spin-orbit 
coupling, and hence we expect a double peak. The peak distance and
their height provides information about the spin-orbit coupling strength, 
and their width points to violations of spin-charge separation.
The ESR spectrum of a MWNT only exhibits a single narrow peak,
whose location, lineshape and linewidth provide 
information about the Rashba-type spin-orbit coupling and 
intrinsic electric fields.
The ESR spectra of SWNTs and MWNTs are therefore fundamentally
different and reflect distinct mechanisms of spin-orbit coupling.

We thank L. Forr{\'o} for discussions. Support by the DFG under the 
Gerhard-Hess program and by the project PICT 99 3-6343 is acknowledged.


\begin{references}
\bibitem{foot2}
For SWNTs, the orbital term can always be neglected.
\end{references}
\end{document}